\documentclass[aps,prl,superscriptaddress,showpacs,twocolumn]{revtex4b5}
\usepackage{epsfig}
\begin{document}
\title{Ruthenocuprates: Intrinsic magnetic multilayers} 
\author{I. \v Zivkovi\'c}
\affiliation{Institute of Physics, P.O.B.304, HR-10 000, Zagreb, Croatia}
\author{Y.Hirai}
\affiliation{Physics Department, University of Wisconsin, Madison, WI 53706, U.S.A.}
\author{B.H.Frazer}
\affiliation{Physics Department, University of Wisconsin, Madison, WI 53706, U.S.A.}
\author{M.Prester}
  \email{prester@ifs.hr}
\affiliation{Institute of Physics, P.O.B.304, HR-10 000, Zagreb, Croatia}
 \author{D.Drobac}
\affiliation{Institute of Physics, P.O.B.304, HR-10 000, Zagreb, Croatia}
\author{D.Ariosa}
\affiliation{Institut de Physique Appliqu\'ee, \'Ecole Polytechnique F\'ed\'erale de Lausanne, CH-1012 Lausanne, Switzerland}
\author{G.Margaritondo}
\affiliation{Institut de Physique Appliqu\'ee, \'Ecole Polytechnique F\'ed\'erale de Lausanne, CH-1012 Lausanne, Switzerland}
\author{I.Felner}
\affiliation{Racah Institute of Physics, Hebrew University, Jerusalem, Israel}
\author{M.Onellion}
  \email{onellion@landau.physics.wisc.edu}
\affiliation{Physics Department, University of Wisconsin, Madison, WI 53706, U.S.A.}

\date{\today} 
%
%
\begin{abstract}
{We report ac susceptibility measurements on polycrystalline samples of $SrRuO_{3}$ and three
ruthenocuprates:  superconducting $RuSr_{2}GdCu_{2}O_{8}$ (Ru-1212), superconducting
$RuSr_{2}Eu_{2-x}Ce_{x}Cu_{2}O_{y}$ (Ru-1222, x=0.5) and nonsuperconducting, insulating
$RuSr_{2}Eu_{2-x}Ce_{x}Cu_{2}O_{z}$ (Ru-1222, x=1.0).  Ac susceptibility of both Ru-1222
compositions exhibit logarithmic time relaxation and `inverted' hysteresis loops.  Ru-1212 samples
exhibit none of these behaviors.  We interpret the magnetic behavior of Ru-1222 in the framework
of weakly coupled magnetic multilayers and argue that superconductivity coexists with
qualitatively different magnetic behaviors.}  
\end{abstract}

\maketitle
%
%
%

The coexistence of superconductivity and long-range magnetic order, and the types of magnetic order
compatible with superconductivity, have long been of interest~\cite{map}. For ruthenocuprate samples,
the precise type of the long-range magnetic order coexisting with superconductivity~\cite{fel1,ber}
remains controversial and intriguing~\cite{lyn,wil}. In this report we find that superconductivity
coexists with qualitatively different magnetic behavior within the ruthenocuprate family. The main
result of this report is that the Ru-1222 ruthenocuprates exhibit unexpected magnetic properties:
By studying ac susceptibility in step-like and continuously sweeping low magnetic fields we report a
pronounced susceptibility switching, logarithmic time relaxations (magnetic aftereffect) and
hysteretic, inverted-in-sense, susceptibility butterfly loops.  Most of these properties have been
individually reported previously in other, nonsuperconducting, magnetic systems, but the Ru-1222
ruthenocuprates exhibits all of these properties.  In marked contrast, superconducting Ru-1212
samples exhibit none of these properties.  We argue that Ru-1222 samples are intrinsic magnetic
multilayers, and provide a model for the magnetic coupling.

Polycrystalline samples were fabricated as published elsewhere~\cite{fel1}. We fabricated $SrRuO_{3}$
polycrystalline samples to serve as a three dimensional itinerant ferromagnet reference material.
Ac susceptibility measurements were performed using a CryoBIND system~\cite{cry}. An ac frequency of
230 Hz and an ac magnetic field amplitude of 0.15 Oe were typically used.  A dc magnetic field
amplitude up to 100 Oe was also used.  Fig.1 shows, from ac susceptibility data, that there are
significant differences in magnetic ordering of Ru-1212 and Ru-1222 sample families. While Ru-1212
is characterized by a single maximum at $T_{N}=133K$, ac susceptibility of Ru-1222 samples exhibit a
peak at $T_{M}$(=85K or 117K, for x=0.5 and x=1.0 samples) and a less pronounced feature in the
temperature range 120-140K (detailed shape depends on x), followed by anomalous susceptibility
behavior extending up to 180K.  Neutron diffraction studies~\cite{lyn} of Ru-1212 indicate that the
sharp ac susceptibility maximum at $T_{N}$ corresponds to the onset of antiferromagnetic long range
order of the Ru moments. Magnetization studies establish that there is also a ferromagnetic
component of magnetic order, usually attributed to weak ferromagnetism of canted Ru
moments~\cite{fel1}. Assigning a precise interpretation to each particular anomaly is not
straightforward~\cite{fus1} and will be, as well as obviously important role of Ce in magnetic
ordering (Fig.1), a subject of separate publication.
In this Letter we focus primarily the temperature range 
of susceptibility maxima of the Ru-1222
\begin{figure}[hb]
\epsfig{file=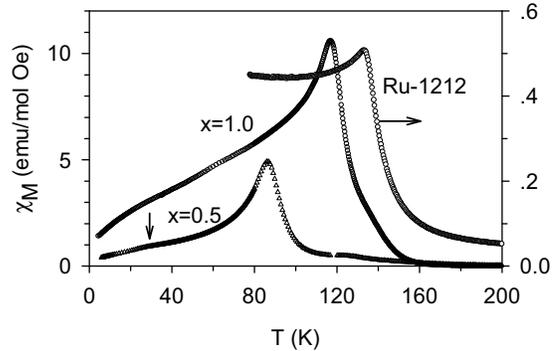,width=7.5cm,clip=}
\caption{Ac susceptibility measurements of Ru-1222 (x=0.5 and x=1.0) and Ru-1212
samples. Note different scales for the two sample types.  Vertical arrow indicates
superconducting transition.}
\end{figure}
\begin{figure}[ht]
\epsfig{file=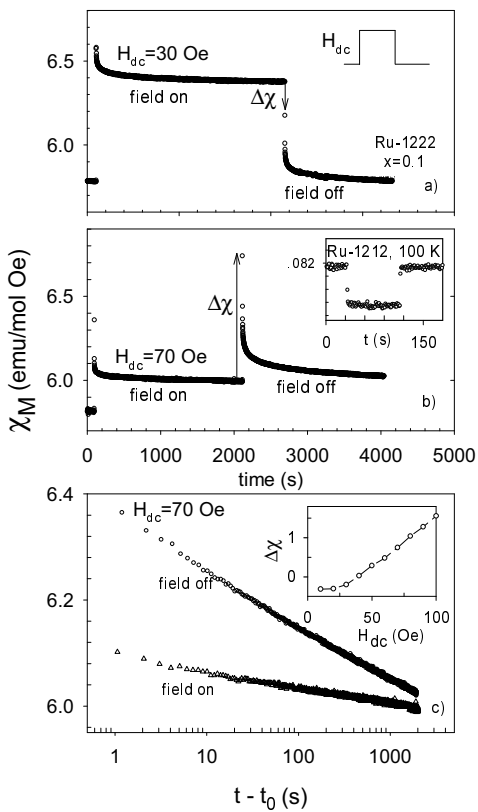,width=6.5cm,clip=}
\caption{a) Ac susceptibility versus time for Ru-1222, x= 1.0 at 80K, with $H_{dc}=0Oe$ initially, then
$H_{dc}=30Oe$, and finally $H_{dc}=0Oe$, shown schematically in the Inset.  ($\Delta \chi$), defined as
change in susceptibility immediately after $H_{dc}$ is switched off, is negative.  Note relaxation of
susceptibility.  b) All conditions the same as in a) except $H_{dc}=70Oe$.  ($\Delta \chi$) is positive
for this value of $H_{dc}$.  Note relaxation of susceptibility.  Inset:  Analogous data for Ru-1212.
Vertical scale is arbitrary units.  Note absence of relaxation.  (c) Ac susceptibility versus logarithm
of time for Ru-1222, x= 1.0, with $H_{dc}=70Oe$ and 80K.  ($t_{0}$) is time at which $H_{dc}$ is either
switched on or off.  Data for both field on and field off conditions included.  Inset:  Overshoot
$\Delta \chi$ (in units of emu/moleOe) at 80K versus $H_{dc}$.  Note change from negative (no
overshoot) to positive (overshoot) at $H_{dc} \approx H_{sf}$.}
\end{figure}
family and report on important new aspects characterizing the complex magnetic order.  The Ru-1222
samples of both compositions (x=0.5, superconductor, and x=1.0, insulator) exhibit qualitatively
similar behavior.  Quantitatively, the effects are more pronounced in x=1.0 sample and Figs.2-4 show
the results for this sample only.  We first report on time relaxation measurements shown in Fig.2.  In
these experiments we monitor ac susceptibility of the zero-field-cooled (ZFC) sample in its response
to the step-like changes ($H_{dc}$) of the dc magnetic field (Fig.2a, inset).  Fig.2 shows the time
dependence of ac susceptibility for two values of $H_{dc}$.  The data indicate:

a) A step-like change of magnetic field causes i) ac susceptibility switch to a new value,
and ii) a sudden transformation of the equilibrium (ZFC) to a metastable magnetic state.
The magnitude of the switch $(\Delta \chi)$ (Fig.  2a, 2b) depends on $H_{dc}$, while its
sign can be positive (for turning the field on) or it can be both negative and positive
(for turning the field off).  The field-induced metastable state is characterized by an ac
susceptibility that relaxes in time, a phenomenon variously attributed to disaccomodation
\cite{mur,kro} or, in wider context of relaxing magnetization, to magnetic
aftereffect~\cite{str}. By contrast, there is no indication of time relaxation in
Ru-1212 samples (Fig.2b, inset);

b) Relaxation is precisely logarithmic in time for both field on and off (Fig.2c), following the
usual form $\chi(t) =\chi_{0} [1- \alpha ln(t-t_{0})]$.  The parameters, $\chi_{0}$ and the
relaxation rate $\alpha$, depend on temperature, $H_{dc}$ , and on the magnetic history (i.e., on
whether the field has been turned on or turned off);

c) For $|H_{dc}|$ above a threshold value of $\approx$ 40 Oe, when $H_{dc}$ is
turned off there is a pronounced `overshoot' phenomenon, i.e., a sizeable positive
$(\Delta \chi)$ (Fig.2b);

d) Applying a rectangular field pulse results in a magnetic state with, surprisingly, {\em
increased} ac susceptibility (Fig.2b).  This state is logarithmically metastable in time.  The slow
relaxation rate indicates the existence of the field-induced excited ac susceptibility for many
time decades.

Next, we report the observation of inverted hysteresis phenomena.  In measuring the
magnetization hysteresis one ramps the applied magnetic field ($H$) from positive to negative
and back and continuously measures the magnetization $M(H)$.  In a similar, `butterfly'
hysteresis technique,\cite{sal} the ac susceptibility $\chi_{ac}(H)$, instead of
magnetization, is measured.  In general, these two hysteresis loops yield similar
information.\cite{fus2} For example, the characteristic maxima in butterfly hystersis (Fig.3a
and inset) correspond to the points in magnetization hysteresis satisfying $M(H)=0$.
Therefore, the positions of butterfly maxima define the coercive field~\cite{sal}
. Fig.3a shows typical butterfly hysteresis data taken for Ru-1222 and Ru-1212 samples.
The data immediately establish that the two types of ruthenocuprates exhibit qualitatively
different responses.  Comparing the Ru-1222 data to the results characterizing the
itinerant bulk ferromagnet $SrRuO_{3}$ (Inset to Fig.3a), there are also pronounced
differences.The most striking difference is the inverted sense of loop circulation for
Ru-1222: the susceptibility signal is consistently larger for the field- decreasing
branch compared to the field- increasing branch.  To our knowledge, this represents the
first observation of the inverted butterfly loops in a bulk magnetic system.  

The numerical integration of the butterfly hystersis is shown in Fig.3b. While the result
for $SrRuO_{3}$ 
\begin{figure}[ht]
\epsfig{file=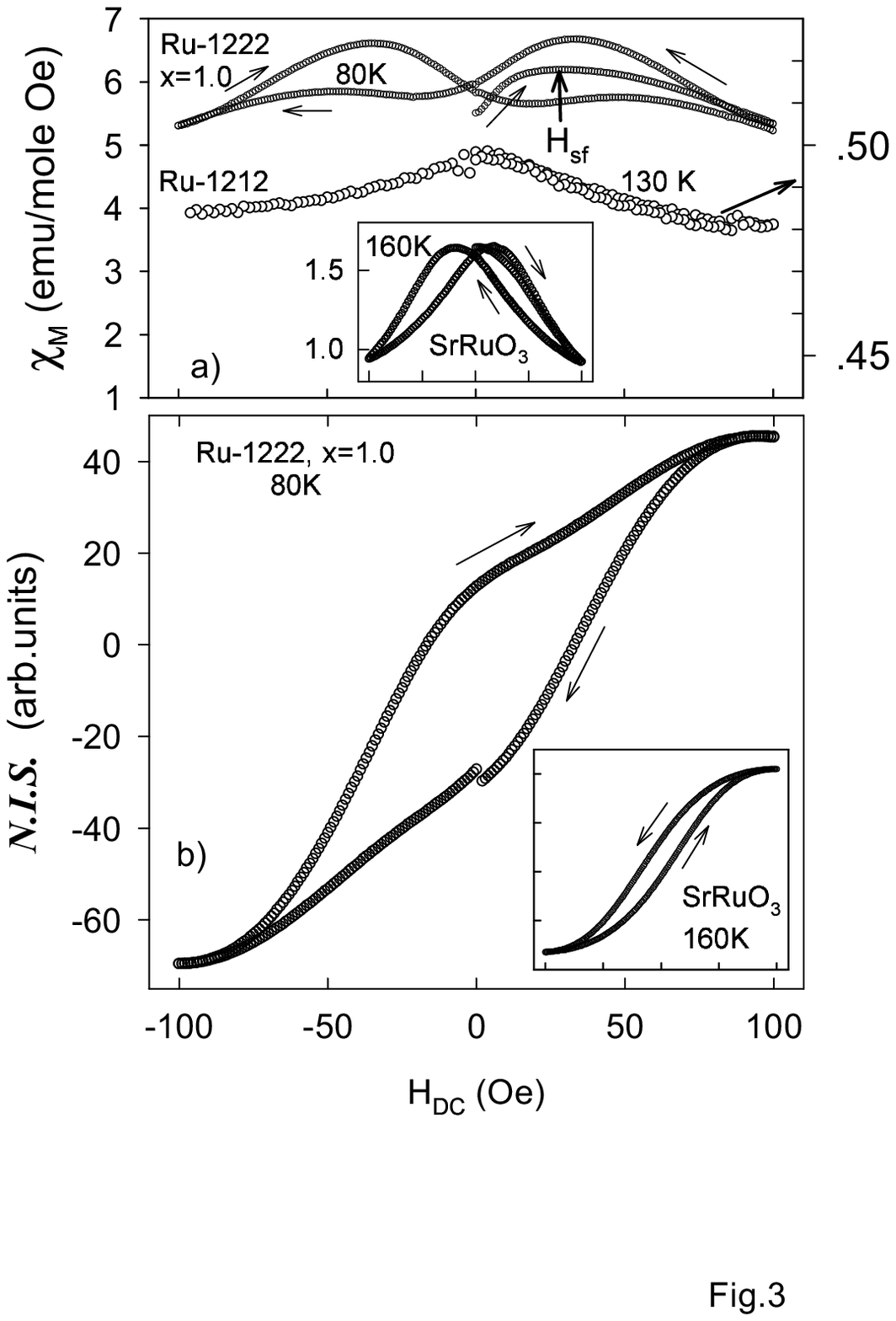,width=7cm,clip=}
\caption[delfsa]{a) Left axis:  `Butterfly' hysteresis for Ru-1222, x= 1.0 at 80K.  Right axis:  Analogous
data for Ru-1212, just below magnetic ordering temperature.  Note just monotonous ac susceptibility
decrease and no hysteresis.  Inset:  Butterfly hystersis for $SrRuO_{3}$ at 160K ($T_{c}= 165K$).
Note the response for increasing and decreasing $H_{dc}$ are opposite to that of Ru-1222.  b)
Numerical integral $NIS$ ($ \equiv \int_{0}^H \chi(h)dh$ ) of butterfly susceptibility shown in a)
versus $H_{dc}$ for Ru-1222.  Note the inverse hysteresis loop.  Inset:  $NIS$ for $SrRuO_{3}$ at
160K.  Note that this hysteresis corresponds, by all means, to the standard ferromagnetic one.}
\end{figure}
exhibits the counter-clockwise pattern of the usual magnetization
hysteresis (Fig.3b inset), the integrated butterfly of the Ru-1222 sample exhibit an
inverted (clockwise) hystersis loop. It is important to note that vibrating sample magnetometer measurements of the same samples, both
$SrRuO_{3}$ and Ru-1222, (data not shown) show the dc magnetization hysteresis loop in the normal
sense of circulation.  Therefore, the inverted hystersis phenomenon represents a unique feature of
the Ru-1222 system, which arises from field-induced metastable states observed in magnetic
aftereffect.  Other noteworthy features of the Ru-1222 butterfly hystersis are i) the presence of a
maximum even in the initial ZFC (virgin) curve, ii) pronounced dependence on the observing time used
to obtain the data, and iii) the presence of the two (instead of only one) maxima per field-ascending
or -descending branch.

As shown in Fig.3a the virgin branch is characterized by a maximum formed at the
characteristic field $H_{sf}$. 
\begin{figure}[ht]
\epsfig{file=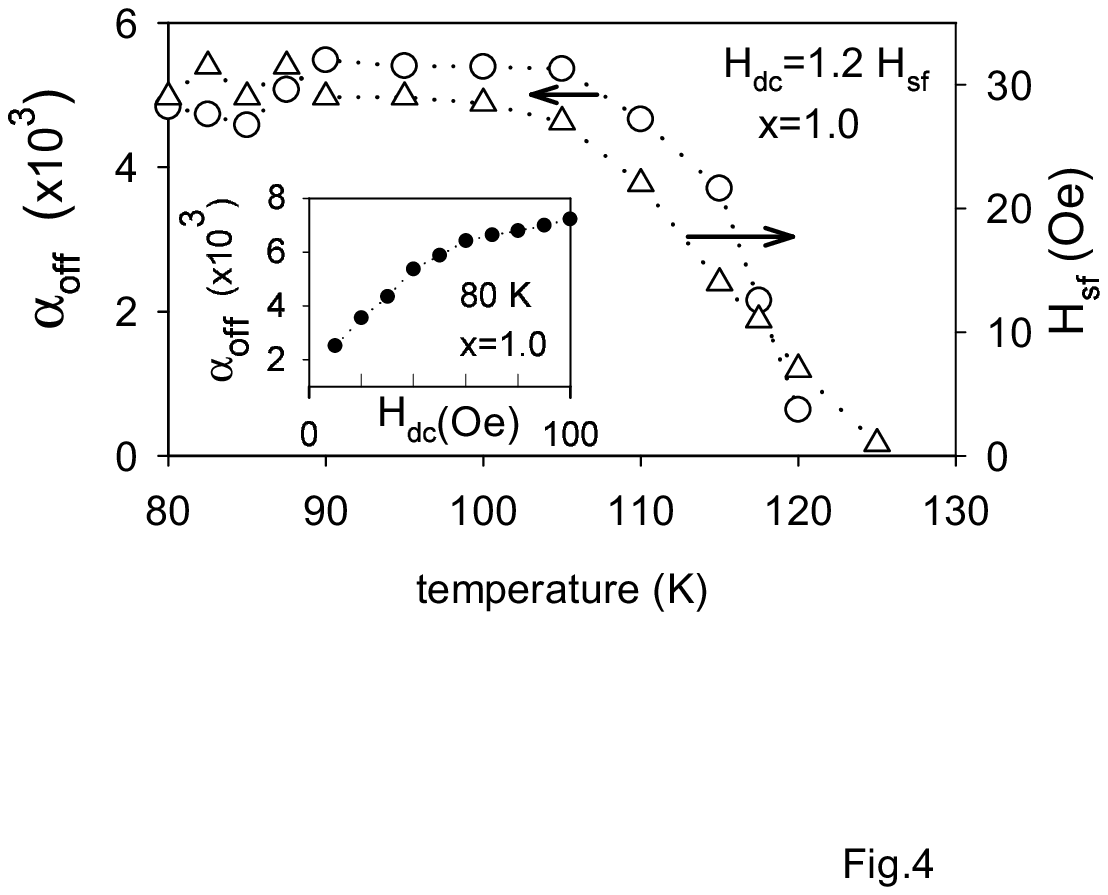,width=7.5cm,clip=}
\caption[delfsa]{Left axis:  $\alpha_{off}$, the relaxation rate constant defined in
text, when $H_{dc}$ is switched off, versus temperature.  Right axis:  The field attributed to spin
flop, $H_{sf}$ (Fig.3a), versus temperature.  Note that both quantities are zero at temperatures
above the susceptibility peak.  Inset:  $\alpha _{off}$, at 80K, versus $H_{dc}$.}
\end{figure}
In ferromagnets the virgin curve typically does not exhibit
a maximum~\cite{fus3} because the remanence, and thus coercive field~\cite{sal}, builds up
only after the first field swing, as seen in the $SrRuO_{3}$ results (Inset to Fig.3a).  The
quantitative size of the butterfly hysteresis loop depends on the observation time, another
indicate that the metastable magnetic states are involved.  Qualitatively, however, over the
range of sweep times we studied- several minutes to one day- the inverted butterfly loops
exhibit the same features.  It is further noteworthy that the field $H_{sf}$ approximately
coincides with the minimum field needed to apply in order to get closed butterfly loops:  if
the range of sweeping field was narrower than $(-H_{sf},+H_{sf})$ no closed loops would be
observed whatsoever.

Figs.2,3 illustrate two interesting phenomena- magnetic logarithmic relaxation (Fig.2) and inverted
hysteretic behavior (Fig.3).  Figure 4 shows that these two phenomena are related in the Ru-1222
system.  Fig.4a illustrates the temperature dependence of $H_{sf}$ (Fig.3) and of $\alpha$, the
logarithmic relaxation rate (Fig.2).  It is noteworthy that the temperature dependence of these two
parameters is virtually identical, indicating that they have a common origin.  Another indication
that the two phenomena are interrelated is shown in the inset of Fig.4a.  Note that $\alpha$
changes rapidly at low fields, but saturates at a $H_{dc}$ field value slightly above $H_{sf}$.
Another quantity that qualitatively changes for fields close to $H_{sf}$ is $\Delta \chi$, shown in
Inset to Fig.2c.  Above $H_{sf}$, $\Delta \chi$ becomes positive and just monotonously increases
with $H_{dc}$.  Again, these qualitative changes in relaxation parameters are connected to the
inverse hysteretic behavior.

Now we interpret our results.  Based on the feature of inverted hysteresis loops and on the
presence and the properties of the field $H_{sf}$, we propose that the magnetism of weakly
antiferromagnetically (AF) coupled magnetic multilayers~\cite{pri} underlies the ordering in
Ru-1222 for the temperature range around $T_{M}$.  Further, based on the time relaxations, we argue
that the weak AF coupling relies primarily on the long range dipole-dipole interaction:  as it is
well-known~\cite{lot} this interaction may be responsible for slow, in particular logarithmic,
relaxation without any additional assumptions.  Our model stems from previous reports of inverted
hystersis loops, limited exclusively to magnetic multilayers \cite{pou} and nanoscale magnetic
films~\cite{cou}. Ref.~\onlinecite{pou} argued for adjacent layers having co-existing magnetic
moments, with AF-like coupling between adjacent layers. Ref.~\onlinecite{pou} also demonstrated
that the inverted hysteresis loop behavior disappears if the AF coupling between adjacent layers is
changed to ferromagnetic coupling.  We argue that the inverted hysteresis loop data (Fig.3)
indicates ferromagnetic intralayer and AF interlayer coupling.  We further argue that the observed
hysteretic behaviors actually originates from spin flop or spin flip processes by which the net
magnetization~\cite{die} of adjacent, weakly antiferromagnetically coupled, $RuO_{2}$ layers
increases.  Compared to bulk antiferromagnets, however, the ruthenocuprate magnetic multilayer spin
flop fields are weaker by several orders of magnitude (which is generally true for atomic-scale
multilayers~\cite{die}), so the hysteresis loops approach saturation for fields of the order of 100
Oe.  We attribute the maxima in the initial ac susceptibility data (Fig.3a) to a spin flop
transition.  This assignment is quite similar to a recent report on AF ordered chain-ladder
compounds~\cite{iso}. Our assignment is further supported by the fact that the dynamical hysteresis
loops close only for applied fields higher than $H_{sf}$.  Previous reports~\cite{fel1} indicate
that the $RuO_{2}$ planes are the main source of magnetic moment for Ru-1222. Consistent with weak
magnetic interactions of our model we argue that only a small interlayer component of canted Ru
moments participates in ferromagnetic intralayer and AF interlayer coupling.

Logarithmic relaxations, considered independently from other observations, could perhaps be
interpreted in framework of domain-wall stabilization (disaccomodation) involving a broad range
of activation energies, as has recently been applied to observations in a
perovskite-manganite~\cite{mur}. However, one of us (IF) and colleagues have performed
temperature-dependent x-ray diffraction studies of Ru-1222; these indicate no structural
change- necessary for disaccommodation- with temperature~\cite{fel3}.  Therefore, this model
favors the active role of dipole-dipole interaction due to its consistency with both the
inferred weak AF coupling \cite{die} and the observed logarithmic relaxations~\cite{lot}.

In summary, we have reported data indicating that Ru-1222 exhibits qualitatively new
magnetic behavior, including magnetic logarithmic relaxation, inverted hysteresis loops,
and metastable magnetic states.  However, superconductivity in the ruthenocuprates is
consistent both with the presence of these metastable magnetic states in Ru-1222 and with
the qualitatively different magnetic behavior of Ru-1212.

We benefited from conversations with Robert Joynt.  Financial support was partially provided by 
the U.S.-Israel Binational Science Foundation, Fonds National Suisse, and EPFL.


\begin{thebibliography}{99}
\bibitem{fus1}There is pronounced deviation from Curie-Weiss (CW) model in the whole
temperature range below 180K.  More generally, the failure of CW model to determine the
magnetic ground state of Ru-1212 demonstrates a questionable validity of this simple mean field
model in interpretation of complex magnetic systems, such as ruthenocuprates.
\bibitem{fus2}The equivalency of butterfly and magnetization hystersis is a simple consequence
of the relationship:  $M(H)=\int_{0}^H \chi(h)dh$ valid - up to an integration constant- in
simple isotropic ferromagnets.  However, in complex magnetic systems replacing $\chi$ (a
tensor) by a scalar is not approved and the relationship does not hold.
\bibitem{fus3}In principle, the observed maximum could be ascribed to a cross over between
reversible and irreversible dynamics of ferromagnetic domains.  However, the idea of
reversibility, at least in its elementary form, is incompatible with logarithmic relaxations
characterizing the whole 0-100 Oe range.  Therefore, it seems that the latter interpretation
cannot be applied in its original transparent form.
\end{thebibliography}
\end{document}